\documentclass[%
 reprint,
 amsmath,amssymb
 aps,
]{revtex4-1}

\usepackage{graphicx}
\usepackage{dcolumn}
\usepackage{bm}
\usepackage{color}
\usepackage{wasysym}
\usepackage{epstopdf}
\usepackage{verbatim}
\usepackage{natbib}
\usepackage{epsfig}
\usepackage{subfigure}

\begin{document}

\preprint{APS/123-QED}

\title{Transporting long-lived quantum spin coherence in a photonic crystal fiber}

\author{Mingjie Xin, Wui Seng Leong, Zilong Chen}
\author{Shau-Yu Lan}%
 \email{sylan@ntu.edu.sg}
\affiliation{%
Division of Physics and Applied Physics, School of Physical and Mathematical Sciences, Nanyang Technological University, Singapore 637371, Singapore
}%




\date{\today}

\begin{abstract}
Confining particles in hollow-core photonic crystal fibers has opened up new prospects to scale up the distance and time over which particles can be made to interact with light. However, maintaining long-lived quantum spin coherence and/or transporting it over macroscopic distances in a waveguide remain challenging. Here, we demonstrate coherent guiding of ground-state superpositions of $^{85}$Rb atoms over a centimeter range and hundreds of milliseconds inside a hollow-core photonic crystal fiber. The decoherence is mainly due to dephasing from residual differential light shift (DLS) from the optical trap and the inhomogeneity of ambient magnetic field. Our experiment establishes an important step towards a versatile platform that can lead to applications in quantum information networks and matter wave circuit for quantum sensing.

\end{abstract}

\pacs{Valid PACS appear here}
\maketitle
Long-lived quantum spin coherence is one of the key requirements in precision measurements and quantum networks \cite{Cro,Kim,Ham}. In atomic interferometric sensors, such as atomic fountain clocks and atom interferometers, atoms are set free in a region of space well-shielded from decoherence sources \cite{Cro}. In quantum memories, the information is encoded in the coherence of spin states, which are normally confined locally in a well-engineered trap to be immune from decoherence \cite{Ham}. In either case, the atoms and coherence are localized within zero or one spatial dimension. To extend the use of quantum coherence into fully three-dimensional coordinates, conveying atoms in a quantum superposition state over a configurable path has been a long-standing goal.

Unlike photons, the quantum coherence of spin states is more susceptible to decoherence from the guiding environment. For neutral atoms, the restoring force used for confining atoms is generally through the coupling to the gradient of optical or magnetic fields \cite{Met}. Coherent guiding of Bose-Einstein condensate using a magnetic trap has been demonstrated on the scale of few hundred micrometers \cite{Ryu}, but the ability to guide and maintain coherence over longer distances is challenging. Although optical conveyor belts in free space have transported a superposition state of atoms over 1 mm range, the Rayleigh range over which a laser beam stays focused limits the guiding distances \cite{Kuhr}. Moreover, the methods above cannot be easily reconfigured to the desired trajectories.

Optical fibre technologies have provided a new platform to extend the scalability and flexibility for guiding atoms. For optical nano-fibres, atoms are trapped a few hundred nanometers away from the surface by the evanescent waves of the guided mode \cite{Sol}. The coherence time of stationary atoms has been demonstrated to few milliseconds \cite{Rei}. Trapping and guiding of atoms in hollow-core photonic crystal fibres \cite{Mar,Ben,Rus} have been demonstrated over centimeter distances \cite{Baj,Oka,Bla,Lan,Xin,Chr,Vor,Hil}; however, the ability to maintain the coherence of quantum superposition states over a large distance is still unclear. A moving atom interferometer has been demonstrated inside a hollow-core photonic crystal fibre, but the coherence time was limited by the dephasing caused by the trapping potential to tens of microseconds \cite{Xin}. Here, we study the quantum coherence of $^{85}$Rb atoms trapped by the fundamental mode of a hollow-core photonic crystal fibre and extend the coherence time to hundreds of milliseconds by mitigating decoherence from the inhomogeneous broadening of DLS. We further demonstrate the transportation of quantum spin coherence over centimeters range.

\begin{figure*}
\includegraphics[scale=0.3]{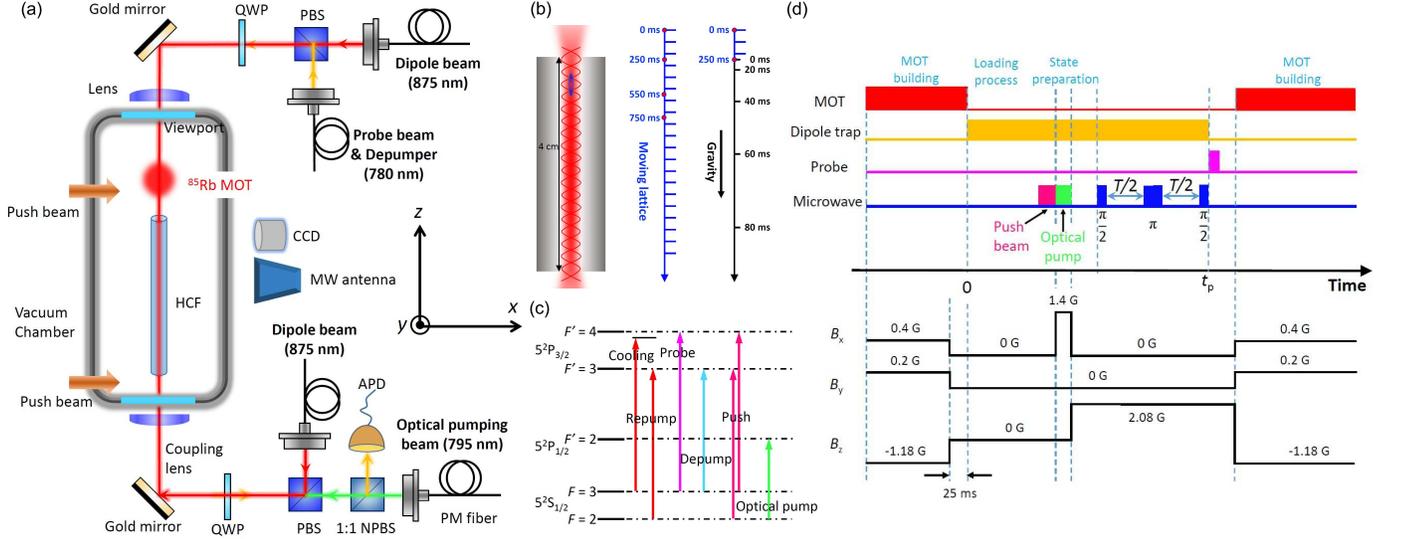}
\caption{\label{fig:epsart} Experimental configuration. (a) Experimental setup. NPBS: non-polarizing beam splitter. PBS: polarising beam splitter. DM: dichroic mirror. QWP: quarter-wave plate. PM fiber: polarisation maintaining fiber. APD: avalanche photodiode. HCF: hollow-core fiber. All the optical components are outside the vacuum chamber. The polarization of the two optical dipole beams is adjusted by the quarter wave plates for differential light shift cancellation. The coupling efficiency of the dipole beams is 65$\%$. (b) Time scales for atoms’ position in the setup. The left scale shows the position of the atoms relative to the fiber in the moving lattice at 2.2 cm/s starting from MOT’s position at $t_{\textrm{p}}$=0 ms. The right scale shows the position of the atoms in the moving lattice at 2.2 cm/s (long ticks) followed by free-falling under gravity while still confined radially by the top dipole beam (short ticks) starting at $t_{\textrm{p}}$=250 ms. For coherence measurements of free-falling atoms, we reset the timing when atoms begin the free-fall. (c) Relevant atomic energy levels. (d) Time sequences. Each experimental cycle includes 400 ms for cooling atoms. $B_{\textrm{x}}$, $B_{\textrm{y}}$, and $B_{\textrm{z}}$ are external magnetic fields for shifting the atomic ensemble and defining quantization axis for atom-light interactions. The $\pi$ pulse is present for spin-echo sequences and absent for Ramsey sequences.}
\end{figure*}

The details of the experimental setup are shown in Figure 1. A 4-cm-long hollow-core photonic crystal fibre from GLOphotonics (PMC-CTiSa-Er-7C) is mounted vertically inside an ultra-high vacuum chamber. The fiber inner core diameter is 63 $\mu$m, and the 1/$e^{2}$ mode field radius $W$ is 22 $\mu$m. The cladding area of the fibre has a hypocycloid-core Kagome lattice structure to inhibit coupling between the fundamental mode and higher order modes \cite{Deb}. It also supports a significant spatial separation between the fundamental mode and the inner wall to minimize the interaction between the surface and atoms \cite{Deb}. The dominant atom-surface interaction in our experiment is the temperature dependent van der Waals potential $U_{\textrm{L}}$=$k_{\textrm{B}}$$T_{\textrm{E}}$$\alpha_{\textrm{0}}$/(4$r^{3}$) , where $k_{\textrm{B}}$ is the Boltzmann constant, $T_{\textrm{E}}$ is the equilibrium temperature of the surface, $\alpha_{\textrm{0}}$ is the static polarizability of the atom, and $r$ is the distance of the atoms from the inner fiber wall \cite{Wol}. This interaction causes atoms at different positions having different resonant frequencies which results in inhomogeneous dephasing when measuring an ensemble of atoms. For $T_{\textrm{E}}$=300 K and $r$=26.5 $\mu$m, the ground state energy shift of $^{85}$Rb is at the level of few mHz which is negligible in our experiment.

A three-dimensional magneto-optical trap (MOT) is aligned about 5 mm above the fibre tip to capture and cool room temperature $^{85}$Rb atoms from background vapor. After sub-Doppler cooling, there are about 10$^{9}$ atoms at a temperature $T_{\textrm{atom}}$=10 $\mu$K. Atoms are released from the MOT at $t_{\textrm{p}}$=0 and loaded into a moving optical lattice formed by a pair of counter-propagating fields in the fiber. The velocity $v$ is determined by $v$= $\delta f\lambda/$2=2.2 cm/s, where $\delta f$=50 kHz is the frequency detuning of the two lattice fields, and $\lambda$= 875 nm is the lattice wavelength. Push beams with 2 ms duration resonant on both the $F$=3 to $F'$=4 and $F$=2 to $F'$=3 transition are sent horizontally to push atoms away from the vicinity of the fiber tip to ensure only atoms inside the fiber participate in the coherence time measurements. The root mean square radius $r_{\textrm{0}}$ of the atomic cloud in the fiber is ($W^{2}$$k_{\textrm{B}}$$T_{\textrm{atom}}$/(2$U$))$^{1/2}$$\sim$ 5 $\mu$m, where $T_{\textrm{atom}}$$\sim$ 90 $\mu$K is the measured atom radial temperature in the lattice, and $U$= 537 $\mu$K is the optical trapping potential.

When atoms are inside the fiber, the detuning between the lattice beams is ramped down to zero to form a stationary lattice. The atoms are then optically pumped to $F$=2, $m$=0 via one 1 ms long intra-fiber $\pi$-polarized optical pump pulse resonant on the D1 line $F$=2 to $F'$=2 transition and one linearly-polarized depump pulse resonant on the D2 line $F$=3 to $F'$=3 transition. We study the coherence of the hyperfine clock states $F$=2, $m$=0 and $F$=3, $m$=0 of stationary $^{85}$Rb atoms using microwave Ramsey $\pi$/2-$T$-$\pi$/2 and spin-echo $\pi$/2-$T$/2-$\pi$-$T$/2-$\pi$/2 sequences, where $T$($T$/2) is the pulse separation time. These sequences have been widely used to identify reversible decoherence mechanisms and, more importantly, to characterize irreversible decoherence mechanisms of atomic coherence \cite{Kuhr2}. For the hyperfine ground states of alkali-metal atoms, different polarizabilities of the states in the optical potential cause different energy shifts which lead to the inhomogeneous DLS of the transition due to the non-uniform transverse profile of the trapping potential.

\begin{figure*}
  \includegraphics[scale=0.8]{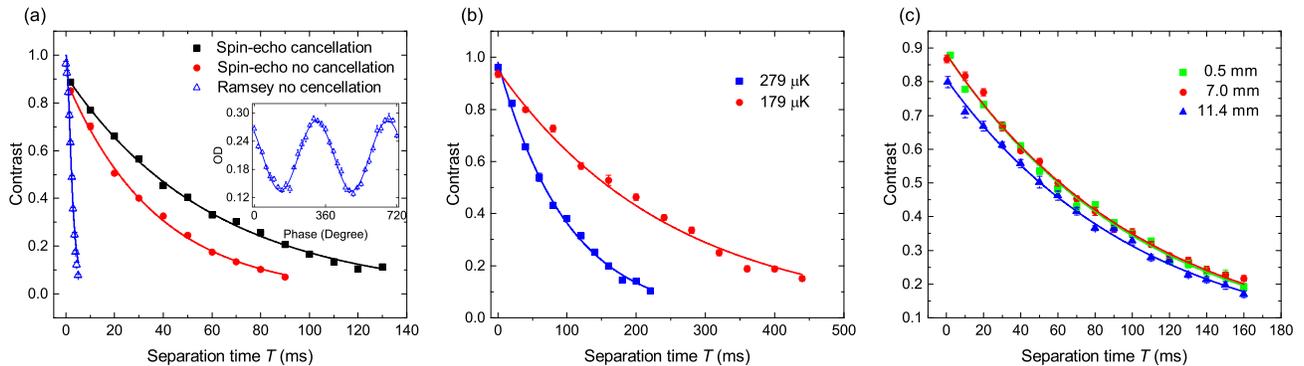}
  \caption{Coherence time measurement of stationary atoms trapped inside the fiber. (a) Contrast versus pulse separation time $T$. The coherence time $T_{\textrm{c}}$ for open triangles, squares, and circles are 3.0(1), 37.2(7), and 60.1(1) ms respectively. The trapping potential of the stationary lattice is 537 $\mu$K. Atoms are transported into the fiber using a moving lattice at 2.2 cm/s for 200 ms and then kept stationary while coherence measurements proceed. Inset: Exemplary Ramsey fringe for pulse separation time $T$=3 ms. The error bars indicate the standard error of 4 experimental runs. The continuous line is a sinusoidal fit to the data from which we extract the contrast $C$. (b) Spin-echo measurement with DLS cancellation of two different trapping potentials 279 $\mu$K (squares) and 179 $\mu$K (circles). The solid lines are fits using the exponential function $A$$\times$exp(-$T$/$T_{\textrm{c}}$), and the 1/$e$ decay time $T_{\textrm{c}}$ are 101(2) and 252(8) ms, respectively. Atoms are transported into the fiber using a moving lattice at 2.2 cm/s for 250 ms and then kept stationary while coherence measurements proceed. (c) Spin echo measurements at different positions of the fiber with DLS cancellation. The 1/$e$ lifetimes for 250, 550, and 750 ms are 106(2), 108(2), and 106(2) ms, respectively which correspond to 0.5, 7.0, and 11.4 mm from the upper tip of the fiber, respectively. The trapping potential of the lattice is 279 $\mu$K.}
\end{figure*}

At the end of a microwave Ramsey or spin-echo sequence, we perform state-selective detection of atoms in $F$=3. This is achieved by measuring the transmission $T_{\textrm{r}}$ of a 3 nW, 50 $\mu$s probe pulse, resonant (far below the saturation power of ~30 nW) on the $F$=3 to $F'$=4 transition, through the fiber. The measured transmission is converted to optical depth OD=-ln($T_{\textrm{r}}$), a quantity directly proportional to atom number in $F$=3 \cite{Xin}. Throughout this paper, we typically measure a maximum OD$\sim$1, which corresponds to about 15000 atoms. By scanning the phase $\phi$ of the last $\pi$/2 microwave pulse, we map out the Ramsey fringe as a sinusoidal oscillation in the OD versus microwave phase $\phi$. Figure 2(a) inset shows an exemplary Ramsey fringe for pulse separation time $T$=3 ms. The contrast C of the Ramsey fringe is obtained from fitting the function $\text{OD}_{\textrm{avg}}$$\times$($C$$\times$sin($\phi$+$\phi_{\textrm{0}}$)+1) to the data with $C$, $\phi_{\textrm{0}}$, $\text{OD}_{\textrm{avg}}$ as fit parameters. The contrast $C$($T$) is then measured as a function of pulse separation time $T$, see Fig. 2(a), and fitted with the function \cite{Kuhr2} (1+($T$/$T_{\textrm{c}}$)$^{2}$)$^{-3/2}$ where $T_{\textrm{c}}$ is the coherence time. Inhomogeneous broadening of the transition from the lattice has limited our Ramsey coherence time $T_{\textrm{c}}$ to 3.0(1) ms as shown in Fig. 2(a).

In principle, the spin-echo sequence can remove this dephasing by reversing the sign of the phase accumulated during the period between the second and third pulse as long as the optical potential experienced by the atoms is time independent \cite{Kuhr2}. With spin-echo, the coherence time $T_{\textrm{c}}$ has improved to 37.2(7) ms as shown in Fig. 2(a) (circles). The spin-echo contrast decay data is fitted with $A$$\times$exp(-$T$/$T_{\textrm{c}}$), where $T_{\textrm{c}}$ and $A$ are the fitting parameters. Time-dependent fluctuations of the potential due to thermal motion of the atoms in the potential, laser power stability or acoustic noise affecting fiber coupling could still cause the residual dephasing. To minimize the DLS, we follow the scheme using a combination of the polarization of the lattice laser beams and an external magnetic field to engineer the polarizability of the $F$=2 and $F$=3 states \cite{Der,Lun,Chi,Dud,Yan,Car,SM}. The cancellation works by using elliptically polarized dipole beams to generate a fictitious magnetic field which causes a Zeeman-like energy shift, called the vector light shift, of the
hyperfine clock states. Figure 2(a) shows the contrast decay of spin-echo interference fringes as a function of the time $T$ with and without DLS cancellation. With spin-echo and with DLS cancellation, the coherence time $T_{\textrm{c}}$ has further improved to 60.1(1) ms as shown in Fig. 2(a) (squares).

\begin{figure}
\includegraphics[scale=0.37]{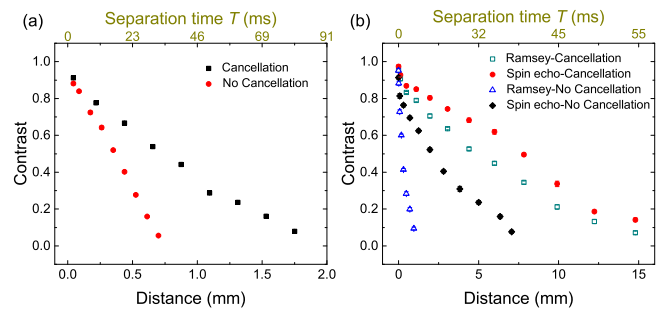}
\caption{\label{fig:epsart} Contrast decay measurement of moving atoms in the fiber. (a) Contrast of spin-echo measurements of atoms in a moving lattice at 2.2 cm/s versus transportation distance with (squares) and without (circles) DLS cancellation. The trapping potential of the moving lattice is 537 $\mu$K. The first $\pi$/2 pulse is applied at $t_{\textrm{p}}$=250 ms. (b) Comparison of contrast measurements of free-falling atoms in an optical dipole trap with different conditions. Atoms are set free to fall under gravity by switching off the bottom dipole beam at $t_{\textrm{p}}$=250 ms while remaining confined radially by the 134 $\mu$K trapping potential from the top dipole beam. The data shows the contrast of the Ramsey fringes with (open squares) and without (open triangles) DLS cancellation and spin-echo sequence with (circles) and without (diamonds) DLS cancellation.}
\end{figure}

As indicated by previous studies, perfect cancellation of the DLS between the hyperfine ground states of alkali-metal atoms in a single trapping beam wavelength does not exist due to hyperpolarizability or non-linear ac Stark shifts \cite{Yan,Car}. The coherence time of our stationary atoms in the waveguide is limited by this residual light shift from the trapping laser. In Figure 2(b), we compare the spin-echo signal at two different trapping potentials 279 $\mu$K and 179 $\mu$K and observe the improvement of the coherence time from 101(2) ms to 252(8) ms. The spin relaxation rate \cite{Kuhr2} calculated at the peak trapping potentials are 0.01 and 0.03 s$^{-1}$ respectively which are negligible for decoherence. We also measure the spin-echo coherence time of the superposition state at different locations along the fiber. The detuning of the lattice beams is ramped from 50 kHz to 0 kHz at $t_{\textrm{p}}$=250 ms, 550 ms, and 750 ms, which corresponds to about 0.5, 11.0, 11.4 mm from the upper tip of the fiber, respectively. As shown in Fig. 2(c), the coherence time of stationary atoms are consistent in three different locations. This demonstrates that the irreversible decoherence is consistent along the fiber.

To demonstrate the transportation of coherent superposition state in the hollow-core fiber, we apply the initial $\pi$/2 microwave pulse in a Ramsey or spin-echo sequence at $t_{\textrm{p}}$=250 ms to put atoms in a superposition state and then let the relative phase of the two states evolve while atoms are in motion. Figure 3(a) shows the contrast of spin-echo sequence when atoms are in the 537 $\mu$K deep, 2.2 cm/s moving optical lattice. The transportation distance is limited by the residual DLS similar to Fig. 2(a). We also switch off the bottom dipole beam at $t_{\textrm{p}}$=250 ms to cause atoms to free-fall under gravity, guided only by the 134 $\mu$K deep top dipole beam. In Figure 3(b), we compare the results of different measurement conditions when atoms are under free fall. Without the DLS cancellation, the spin-echo coherence time greatly exceeds the Ramsey coherence time. With the cancellation, however, these two sequences show similar decoherence trend. We could still observe the contrast after 55 ms of the Ramsey pulse separation time which corresponds to the transportation of 1.5 cm.
\begin{figure}
\includegraphics[scale=0.4]{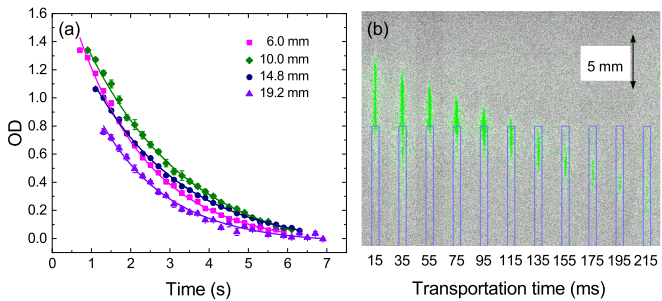}
\caption{\label{fig:epsart} Atoms in the fiber. (a) Lifetime measurements of atoms trapped in a stationary lattice. (b) Fluorescence images of atoms at different times during transportation by a moving lattice at 4.4 cm/s. The atoms are probed with resonant light on $F$=3 to $F'$=4 cycling transition, and scattered photons are imaged from the side of the fiber onto a CCD camera. The fiber location is indicated by the blue rectangle. The 5 mm scale bar is calibrated using calculated lattice velocity and the centroid position of the atoms versus time. Atoms that are not following the moving lattice potential fall faster than atoms in the moving lattice, which can be seen in the image of transportation time 35 ms.}
\end{figure}
Figure 4(a) shows the lifetime of atoms inside the stationary lattice at different positions of the fiber. Atoms are transported by a moving lattice at 2.2 cm/s into the fiber and held at $t_{\textrm{p}}$= 500, 700, 900, and 1100 ms for lifetime measurement, which correspond to 6.0, 10.0, 14.8, and 19.2 mm from the upper tip of the fiber, respectively. The lattice depth is 537 $\mu$K. We measure the decay of the atom number as a function of time and fit with an exponential decay function. The 1/$e$ lifetimes of the atoms are 1.87(4), 2.59(7), 2.33(5), and 1.74(7) s, respectively. The minimum lifetime of 1.7 s indicates the background pressure inside the fiber is better than 1$\times$10$^{-8}$ torr \cite{Oka} and sets the upper bound of our coherence time measurements. Figure 4(b) shows the fluorescence images of atoms at different times during transportation by a moving lattice at 4.4 cm/s.

The long coherence time of stationary atoms in our experiment can be used to build a fiber-based atomic clock \cite{Ili}. The decoherence is due to the residual DLS from the 11 mG inhomogeneous magnetic field gradient around the fiber holder over our sample size of 5 mm \cite{SM} which translates into few Hz of residual DLS at 537 $\mu$K trapping potential. This can be improved by removing the magnetic field gradient or exciting ring-shaped higher-order modes of the fiber with blue detuned dipole laser frequency \cite{Dav}. A factor of 10 improvements can be expected when we lower the trapping potential by a factor of 10 while keeping the same ratio of atoms’ temperature to potential depth.

The source of decoherence on moving atoms is most likely due to the inhomogeneity of the local magnetic field along the atoms trajectories. The 11 mG magnetic field gradient corresponds to tens of Hz inhomogeneous broadening on the clock states through second-order Zeeman shift which is cancelled for stationary atoms through spin-echo sequence. The inhomogeneity of the magnetic field also reflects in the fluctuation of the data in Fig. 3. This can be easily improved by an order of magnitude by removing magnetic substance on our fiber holder. This will translate into more than 1-meter transportation of the quantum state in the free-falling case. We observe no dependence of the coherence time on the external magnetic field current noise as well as the gradient when we double the magnetic field. The coherence time is also independent of the trapping potential depth used in the free-falling sequence.

The distance of the hollow-core fiber enhanced quantum state transportation demonstrated in this work is already an order of magnitude longer than the free space scheme \cite{Kuhr}. In the future, we plan to study the coherence of external states of atoms (matter waves) inside the fiber. If long matter wave coherence times approaching the internal state coherence times demonstrated here can be achieved, then the prospects for large enclosed area matter wave interferometer is promising. Considering a fiber with 5 cm bending radius at 3 dB bending loss \cite{Alh}, a matter wave circuit with 30 cm perimeter for quantum sensing purposes can be envisioned \cite{Bar}.

This work is supported by Singapore National Research Foundation under Grant No. NRFF2013-12, Nanyang Technological University under start-up grants, and Singapore Ministry of Education under Grants No. Tier 1 RG107/17.


\nocite{*}

\bibliography{apssamp}

\end{document}


\preprint{APS/123-QED}

\title{Supplemental Material
 for "Transporting long-lived quantum spin coherence in a photonic crystal fiber"}

\author{Mingjie Xin, Wui Seng Leong, Zilong Chen}
\author{Shau-Yu Lan}%
 \email{sylan@ntu.edu.sg}
\affiliation{%
Division of Physics and Applied Physics, School of Physical and Mathematical Sciences, Nanyang Technological University, Singapore 637371, Singapore
}%




\date{\today}

\pacs{Valid PACS appear here}
\maketitle

\section{Timing sequences and detection method}
In Figure 1(d) of the main text, we show the relevant timing sequences for the experiment. Atoms are captured and cooled from the room temperature vapour for about 400 ms with the sub-Doppler cooling process during the last 25 ms. Using the time-of-flight measurement, we measure the temperature of the atoms to be 10 $\mu$K before loading into the moving lattice. Push beams on the top and bottom of the fiber tips are switched on for 2 ms before the coherence time measurement to remove atoms outside the fibre. We also perform optical pumping for 1 ms to initialize atoms in the $F$=2, $m$=0 state. The quantization axis is aligned along the polarization of the optical pumping beam in the $x$-direction during optical pumping and aligned along $z$-direction during coherence measurements. The $\pi$/2 pulses for the Ramsey and spin-echo measurements are 7.5 $\mu$s, and the frequency of the microwave is phase locked to a 10 MHz clock from SRS FS725 Rubidium Frequency Standard. The number of atoms is estimated by measuring the transmission $T_{\textrm{r}}$ of a 3 nW probe pulse resonant on the $F$=3 to $F'$=4 transition with 50 $\mu$s duration and using the relation of optical depth OD=-ln($T_{\textrm{r}}$)=$n$$\sigma$$L$, where $n$ is the number density of atoms, $\sigma$ is the scattering cross-section of atoms, and $L$ is the length of the atomic ensemble \cite{Xin}. Throughout the experiment, OD=1 corresponds to about 15000 atoms. We estimate that the MOT has about 5$\times$10$^{8}$ atoms and only about 15000 atoms are loaded into the fiber. The low loading efficiency is due to the low lattice trapping potential at the center of the MOT because the 13 mrad divergence of the lattice beams from the fiber tip leads to ~7x reduction in trapping potential at MOT than in the fiber. The MOT also has poor geometrical overlap with the diverging lattice beams.

\section{Magnetic field gradient measurement}
The source of decoherence on the moving atoms is mainly due to the magnetic field gradient. To estimate the magnetic field gradient of our setup, we measure the linewidth of magnetic field sensitive $F$=2, $m$=1 to $F’$=3, $m$=2 transition using microwave spectroscopy and compare it with the magnetic field insensitive $F$=2, $m$=0 to $F’$=3, $m$=0 transition. Top of Fig. 1 shows the measurement of the transition linewidth of 1.55 kHz for magnetic field insensitive transition. Assuming the broadening on the insensitive transition is mainly due to the power broadening of the microwave, we fit the Voigt function to the sensitive transition measurements (bottom) with the Lorentz width fixed to 1.55 kHz. The fit result shows a gradient of 11 mG over the sample size of about 5 mm. This gradient translates into the broadening of about 59 Hz on the $F$=2, $m$=0 to $F’$=3, $m$=0 transition due to the second order Zeeman shift at the magnetic field of 2.08 G. To confirm the broadening is coming from the spatial magnetic field gradient instead of the current noise of our coil, we also measure the gradient at different external magnetic fields and observe no dependence on the current noise.
\begin{figure}
  \includegraphics[scale=0.38]{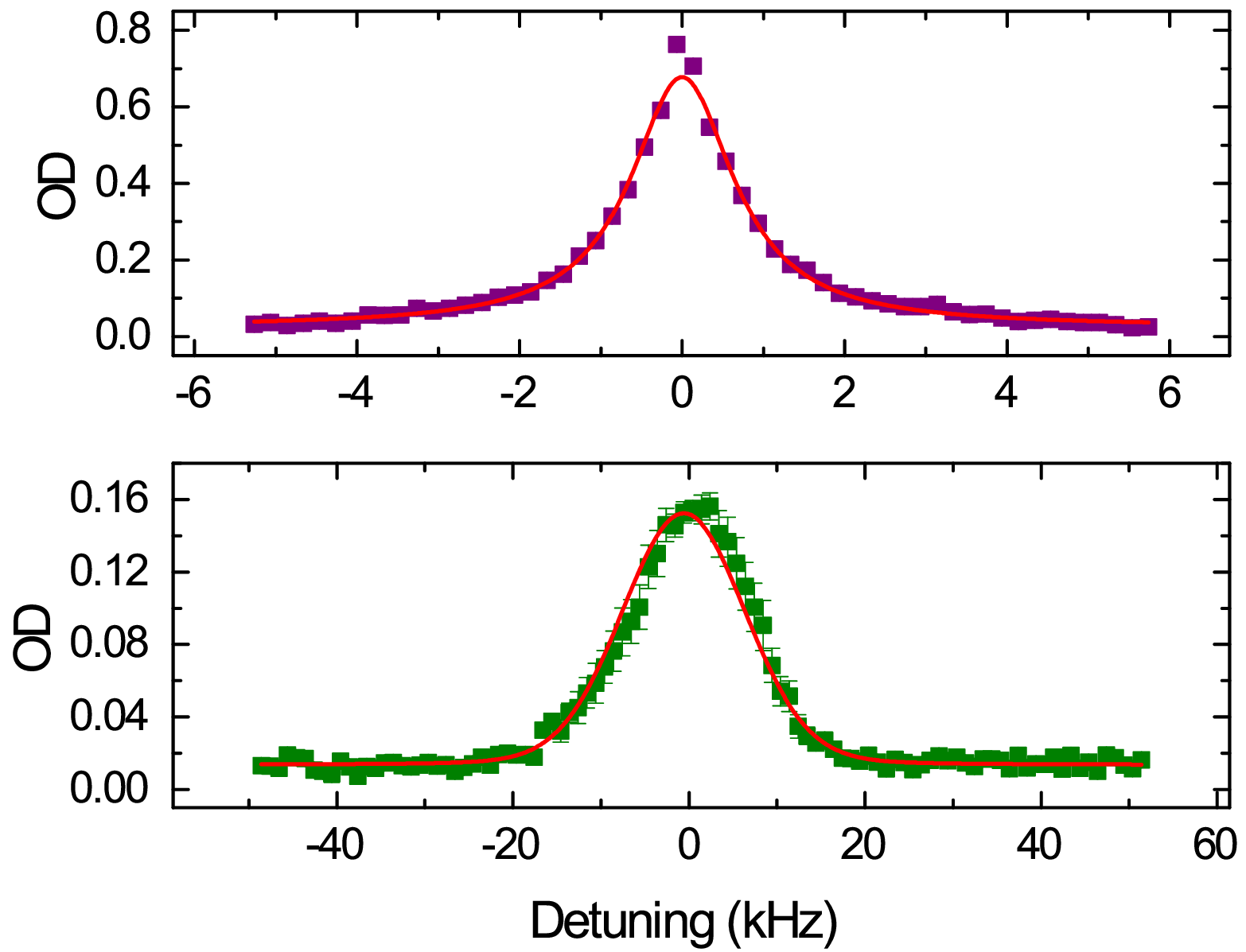}
  \caption{Zeeman spectroscopy of atoms trapped in a linear polarized stationary optical lattice. An external 2.08 G magnetic field is applied to break the Zeeman degeneracy. Top figure is the transition from $F$=2, $m$=0 to $F’$=3, $m$=0 and the bottom figure is $F$=2, $m$=1 to $F’$=3, $m$=2. The top figure is fitted to a Lorentz function with FWHM 1.55(3) kHz. The bottom figure is fitted to the Voigt function with Gaussian width of 15.7(4) kHz (bottom), where the Lorentz width is fixed to 1.55 kHz.}
\end{figure}

\section{Differential light shift cancellation}
A so-called “magic wavelength” optical dipole trap for zeroing the differential polarizability of the two hyperfine ground states of alkali-metal atoms does not exist because of their same electronic structure \cite{Der,Lun,Dud,Yan,Car}. Although the cancellation of the DLS of the hyperfine ground states using vector light shift and additional magnetic field has been proposed and demonstrated, the cancellation is not perfect due to the hyperpolarizability \cite{Yan,Car}. Despite these facts, we still follow ref.\cite{Yan} to cancel the differential light shift partially. The partial cancellation works by using elliptically polarized dipole beams to generate a fictitious magnetic field which causes a Zeeman-like energy shift, called the vector light shift, of the hyperfine clock states. The hyperfine clock states acquire sensitivity to the fictitious magnetic field through a bias magnetic field. The magnitude and sign of the differential vector light shift can be made opposite to that of the differential scalar light shift by adjusting the ellipticity of the dipole beams, and hence cancellation can be obtained up to first order differential scalar light shifts. More precisely, the DLS of two hyperfine ground states in a nonzero magnetic field can be calculated as $\delta$$\nu$($B$,$U$) =$\beta_{1}$$U$+$\beta_{2}$$P$$B$$U$+$\beta_{4}$$P^{2}$$U^{2}$, where $B$ is the bias magnetic field, $U$ is the trapping potential, $P$ is the polarization of the trapping beams ($P$=1 for circular polarization and $P$=0 for linear polarization), and $\beta_{1,2,4}$ are the light shift coefficients. $\beta_{1}$ is defined as the ratio of the differential polarizability of two hyperfine ground states and the ground state polarizability [$\alpha_{\textrm{F=3}}$($\omega$)-$\alpha_{\textrm{F=2}}$($\omega$)]/$\alpha_{\textrm{5s}}$($\omega$). $\beta_{2}$ is defined as -4$\mu_{\textrm{B}}$$\alpha^{v}$($\omega$)/[$h$$f$$\alpha_{\textrm{5s}}$($\omega$)], where $\mu_{\textrm{B}}$ is the Bohr magneton, $\alpha^{v}$($\omega$) is the vector polarizability, $h$ is the Planck constant, and $f$ is the frequency of hyperfine splitting.$\beta_{4}$ can be expressed as $f$($h$$\beta_{2}$)$^{2}$/(8$\mu_{\textrm{B}}^{2}$). The light shift coefficients used in our experiments are $\beta_{1}$=7.86$\times$10$^{-5}$, $\beta_{2}$=-1.13$\times$10$^{-4}$ G$^{-1}$, and $\beta_{4}$=2.48$\times$10$^{-12}$ Hz$^{-1}$. The magnetic field $B$ is set at 2.08 G, and we optimize the coherence time by changing the polarization $P$ of the optical dipole beams using quarter-wave plates.


\nocite{*}

\bibliography{apssamp}